\documentclass[superscriptaddress,aps, prd, reprint, nofootinbib]{revtex4-2}
\usepackage{amsmath}
\usepackage{amssymb}
\usepackage{amsbsy, amstext, amsthm}
\usepackage{dcolumn}
\usepackage{bm}
\usepackage{blkarray}
\usepackage{slashed}          
\usepackage{mathtools}                   
\usepackage{hyperref,graphicx}           
\usepackage{url}
\usepackage{endnotes}
\usepackage{caption}
\usepackage{subcaption}
\usepackage{multirow}
\usepackage{units}
\usepackage{physics}
\usepackage[usenames,dvipsnames,svgnames,table]{xcolor}
\usepackage{graphicx}
\usepackage{tikz}
\usepackage{siunitx}[=v2]
\usepackage[all]{hypcap}
\usepackage{cleveref}
\usepackage[mathlines]{lineno}
\usepackage{tikz-3dplot}
\usepackage{color}
\usepackage{multirow}
\usepackage{float}

\numberwithin{equation}{section}
\definecolor{darkblue}{rgb}{0,0,0.5}
\definecolor{darkgreen}{rgb}{0,0.5,0}

\newcommand{\cmt}[1]{}

\newcommand{\beq}{\begin{eqnarray}}
\newcommand{\eeq}{\end{eqnarray}}
\newcommand{\beqa}{\begin{eqnarray}}
\newcommand{\eeqa}{\end{eqnarray}}

\let\deltafunc\delta
\DeclareDocumentCommand\delta{}{\trigbraces{\deltafunc}}

\graphicspath{{./Figures/}}

\begin{document}
\title{Pulse and Polarization Structures in Axion-Converted X-rays from Pulsars}
\date{\today}

\author{JiJi Fan}
\affiliation{Department of Physics \& Brown Theoretical Physics Center, Brown University, Providence, RI, 02912, USA}
\email{jiji_fan@brown.edu}

\author{Lingfeng Li}
\affiliation{Department of Physics, Brown University, Providence, RI, 02912, USA}
\email{l.f.li165@gmail.com}

\author{Chen Sun}
\affiliation{Abdus Salam International Centre for Theoretical Physics, Strada Costiera 11, 34151, Trieste, Italy} 
\email{csun@ictp.it}

\begin{abstract}
Neutron stars (NS's) with their strong magnetic fields and hot dense cores could be powerful probes of axions, a classic benchmark of feebly-coupled new particles, through abundant production of axions with the axion-nucleon coupling and subsequent conversion into X-rays due to the axion-photon coupling. In this article, we point out that the pulsation structures in both the intensity and polarization of X-rays from NS's could provide us additional information about axions and their couplings. We develop new analytical formalisms of pulsation-polarization structure applicable to a wide range of NS's in the axion scenario and argue that they hold in complicated astrophysical environments. As a case study, we apply our formalism to a representative X-ray Dim Isolated Neutron Star, RX J1856.6-3754, with an unexpected hard X-ray excess which might be axion-induced. We show with an updated fit that the axion explanation is compatible with both the intensity and pulsation data available, and combining the pulsation data does not shift the posterior by more than $1\,\sigma$. Yet, the preferred parameter space is close to being excluded by other astrophysical constraints. With a 75\% reduction of the uncertainties in the pulsation data, we could potentially draw a definite conclusion on the axion-induced X-rays at more than $3\,\sigma$ level. 
\end{abstract}

\maketitle

\section{Introduction}

An axion, a pseudo-scalar field with a discrete shift symmetry, could have a wide range of phenomenological and cosmological applications. Its most notable applications include solving the strong CP problem~\cite{Peccei:1977hh, Peccei:1977ur,Weinberg:1977ma,Wilczek:1977pj}, as well as serving as a classic cold dark matter candidate~\cite{Preskill:1982cy,Dine:1982ah,Abbott:1982af}. Though the axion scenario was proposed more than 40 years ago, it continues to serve as one of the most motivated hypothetical feebly-coupled particles beyond the Standard Model of particle physics, with heightening interests in all kinds of search strategies. Among the astrophysical probes, neutron stars (NS's) with high nucleon densities could be powerful factories for axions, in the presence of axion couplings to nucleons~\cite{Brinkmann:1988vi,Weber:1999qn,Sedrakian:2006mq,Page:2013hxa,Sedrakian:2015krq}. The axions produced could subsequently convert into hard X-rays in the strong magnetic fields around the NS's, if axions also couple to photons. Such a combined process has been studied intensively in the literature, for different types of NS's~\cite{Morris:1984iz,PhysRevD.37.1237,Fortin:2018ehg,Buschmann:2019pfp}, and already gives us highly nontrivial constraints and even potential hints of axions.  

In this article, we study new observables beyond the overall X-ray flux from rotating NS's, the pulsars, which could potentially provide further valuable information about axions and their couplings.\cite{ns-pulsar}
Since the NS magnetic field configuration is not axisymmetric around its spin axis generically, the axion conversion rate depends on the phase of rotation, giving rise to interesting pulse structures in the converted X-ray intensity. In addition, the X-ray photons converted from the axion are nearly 100\% linearly polarized with the associated electric field aligned along the local NS magnetic field. As a result, the polarization signal is also phase-dependent as the magnetic field configuration changes with the pulsar rotation. 

For an isolated NS without a strong accretion process, the axion-to-photon conversion happens dominantly in regions far from the NS surface. Thus, unlike the astrophysical sources of X-rays, which are usually located 
close to the surfaces of NS's (e.g. local hot spots \cite{Yoneyama:2018dnh, DeGrandis:2022kfe}), 
the predictions of the polarization and pulsation signals in the axion scenario are not susceptible to complications in modeling the NS surface and atmosphere. In addition, the higher moments of the magnetic field beyond the dipole one fade away much faster in the effective axion-photon conversion region, and have much less impact on the axion-induced signals than on astrophysically originated X-rays. Consequently, the results obtained in this article are applicable to a wide range of NS's, including X-ray Dim Isolated Neutron Stars (XDINS) and magnetars. The joint pulse-polarization structure from axion-induced X-ray flux can help discern potential axion signals from astrophysical backgrounds. We note in passing that the opposite photon-to-axion conversion can potentially induce interesting polarization signals (see \textit{e.g.} \cite{Perna:2012wn,Song:2024rru} and references therein), although such studies focus on the surface photon spectrum and require careful modeling of the astrophysical environment and standard model contribution to the polarization aside from the axion-induced one.

Taking into account the pulse structure, we revisit the axion explanation for the keV X-ray excess of XDINS~\cite{Buschmann:2019pfp}, taking RX J1856.6-3754 (J1856) as an example. We show that the pulse structure from the axion explanation of the X-ray excess is compatible with the current spectrum and pulsation data. However, the preferred parameter space is on the verge of being excluded by other astrophysical bounds. We will also demonstrate how a solid conclusion on the axion explanation could be achieved with reduced uncertainties in the pulsation data, which are feasible in the near-future observations.

\section{Axion Production and Conversion in the keV-band}
\label{sec:pol}
The hot dense core of a NS mainly consists of nucleons and becomes a significant source of axions with energy in the keV-band in the presence of an axion-nucleon coupling. Here we follow the conventional way of parametrizing coupling of an axion, $a$, to a fermion $f$ with mass $m_f$, as $\mathcal{L}\supset g_{af} \partial^\mu a \bar{f}\gamma_\mu \gamma^5 f/(2 m_f)$, where the coefficient $g_{af}$ is dimensionless. We set the axion coupling to a nucleon (denoted by a subscript $N$), either a proton ($p$) or a neutron ($n$) to be the same $g_{aN}=g_{ap}=g_{an}$. The axion production mechanism in the NS core is dominated by nucleon bremsstrahlung, while the Cooper-pair breaking may become important for certain temperature ranges~\cite{em-around-pulsars}.
Both mechanisms are subject to large systematic uncertainties from high-density nuclear physics and NS interior conditions. As a leading approximation, the axion emission from the NS core is spherically symmetric. For further discussions, see~\cite{Fortin:2021sst}.

The produced axion could also couple to the electromagnetic field through the operator $-(g_{a\gamma}/4) a F_{\mu\nu}\tilde{F}^{\mu\nu}$, where $g_{a\gamma}$ is the dimensionful coupling with energy dimension $-1$ and $F (\tilde{F})$ the (dual) electromagnetic field strength. This operator leads to $a \bold{E} \cdot \bold{B}$ with $\bold{E} (\bold{B})$ being the electric (magnetic) field. It determines that the photons converted from axions are always polarized in the direction of the local background magnetic field. In other words, X-rays from the conversion are always in the so-called ordinary mode ($O$-mode) parallel to the $\bold{k}-\bold{B}$ plane with $\bold{k}$ the photon wave vector. 

\begin{figure}
\includegraphics[width=8.5 cm]{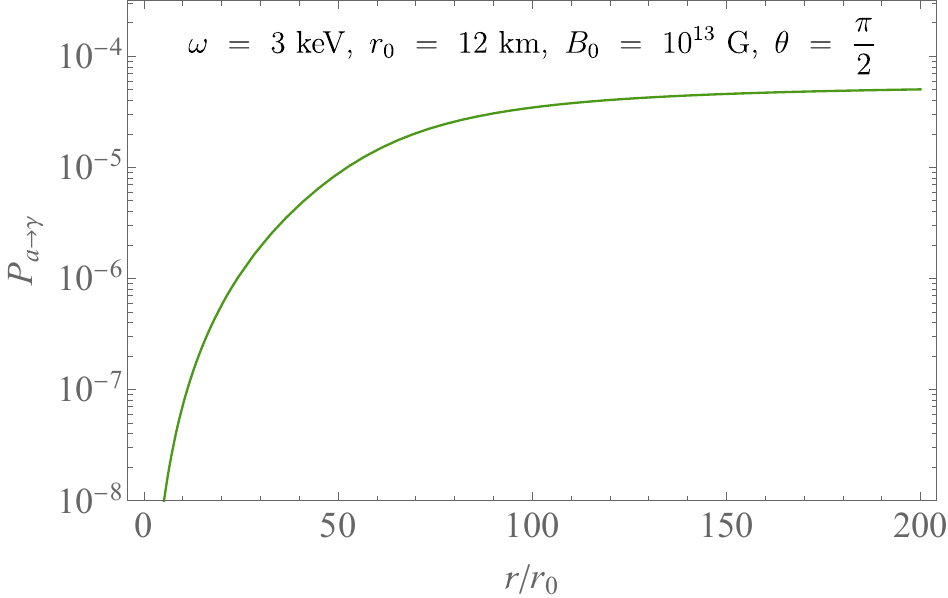} 
\caption{Probability of axions converting to photons ($P_{a\to \gamma}$) as a function of $r/r_0$. It approaches half of its asymptotic value at $\sim 85\,r_0$, distant from the surface, for the benchmark shown. 
}
\label{fig:conprob}
\end{figure}

The axion-photon conversion probability could be computed from the leading-order solution to the time-dependent perturbation theory in quantum mechanics~\cite{PhysRevD.37.1237}. 
A benchmark conversion probability as a function of radial distance $r$ in the unit of $r_0$, the NS radius is shown in Fig.~\ref{fig:conprob}. We define the axion-photon conversion radius, $r_{\rm con}$, to be the scale at which the conversion probability is half of the asymptotic value,
 which is given by
\begin{align}
\label{eq:rcondefinition}
\frac{r_{\rm con}}{r_0}
  & = 85  \;
    \left(\frac{\omega}{3 \, {\rm keV}}\right)^{\frac{1}{5}} 
    \left ( \frac{B_0}{10^{13}\,\textrm{G}}\right )^{\frac{2}{5}}
    \left ( \frac{r_0}{12\,\textrm{km}}\right )^{\frac{1}{5}}
    (\sin\theta)^{\frac{2}{5}}\,    ,
\end{align}
where $\omega$ is the photon energy, $r_0 \approx 12$ km the typical NS radius~\cite{Burgio:2021vgk}, and $\theta$ the angle between the direction of axion(photon) propagation and the local magnetic field. In the above equation, we only consider the dipole component of the magnetic field which scales as $B_0 (r/r_0)^{-3}$ with $B_0$ the magnitude of the dipole field at the surface.
From Fig.~\ref{fig:conprob}, one could see that for axions propagating in the NS magnetosphere, the conversion probability is negligible at small radii as the large effective photon mass induced by vacuum polarization suppresses the mixing. The conversion happens dominantly at larger radii. Thus, higher-order moments of the magnetic field, which vanish faster with the radius, have negligible effects on the conversion rate. 

Note that $r_{\rm con}$ is independent of the axion mass $m_a$ as long as $m_a \lesssim (\omega^2 m_e^2/(\alpha r_0^3 B_0 \sin\theta))^{\frac{1}{5}}~$\cite{Fortin:2021sst}, with $m_e$ the electron mass and $\alpha$ the fine-structure constant. This remains true for most NS's except when $\sin \theta \ll 1$ near the magnetic poles. Yet the axion-photon conversion around the pole is suppressed $\propto |\sin \theta|^{2/5}$. 
Moreover, since the conversion dominantly happens far away from the NS, predicted properties of converted X-rays are robust against astrophysical factors effective close to the NS. These include vacuum resonance in the thin atmosphere, which is only about $\mathcal{O}$(cm) thick~\cite{Lai:2003nd}, and photon bending due to general relativity.
On the other hand, astrophysical sources of hard X-rays such as local hot spots~\cite{Yoneyama:2018dnh, DeGrandis:2022kfe} are subject to all these effects. As shown in the NS literature (summarized in~\cite{Caiazzo_2019}), these complications have competing effects on the polarization and usually predict a mixture of the $O$-mode and $X$-mode (orthogonal to the $\bold{k}-\bold{B}$ plane) with an energy-dependent polarization fraction, in contrast to the clean polarization signal in the axion scenario: a dominant $O$-mode with a polarization fraction $\sim 1$ at a given phase.

\section{Pulses in intensity and polarization}
\label{sec:pulse}
The axion-converted X-ray photons have specific pulse structures in both the intensity and polarization, which are under-explored.

\begin{figure}[h!]
\centering
\includegraphics[width=8cm]{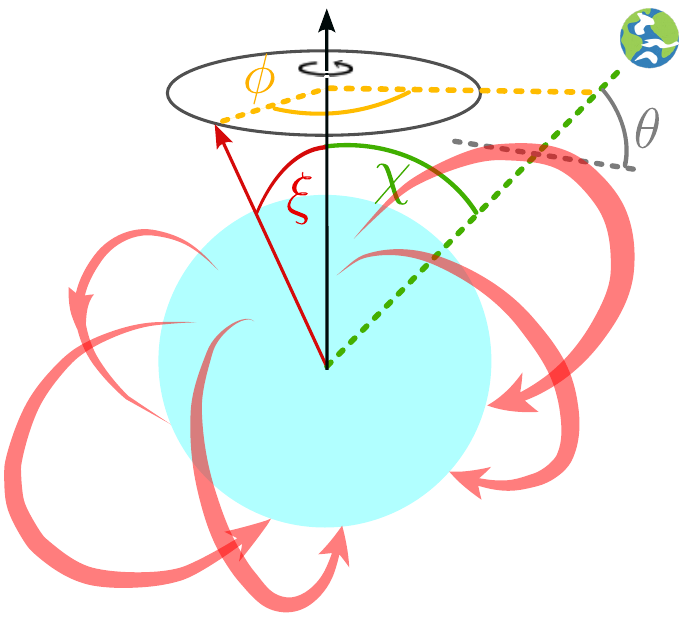}
\caption{Schematic of pulsar field configuration and definition of the $\chi$, $\xi$ and $\phi$ angles. }
\label{fig:angle}
\end{figure}

\begin{figure}[h!]
\centering
\includegraphics[width=8.5cm]{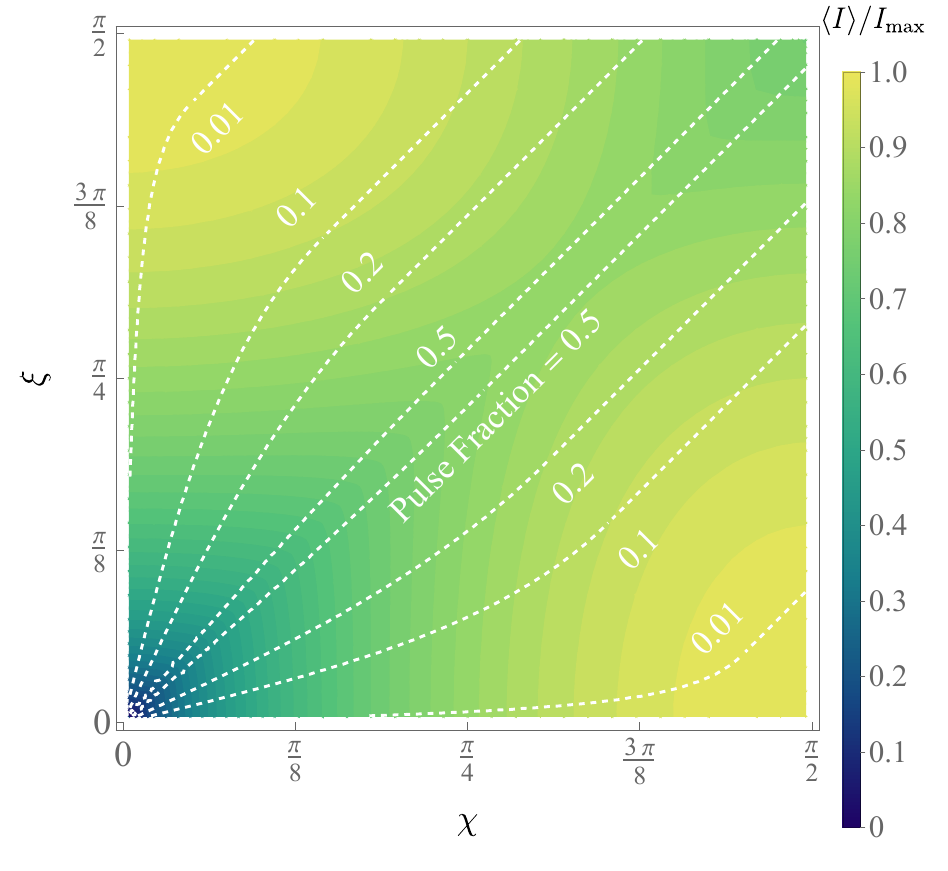}
\caption{The phase-averaged attenuation factor of the photon flux from axion conversions compared to $I_{\rm max}$, color-coded. Both the magnetic dipole approximation and the central LOS approximations are adopted. The white contours show the pulse fraction.
} \label{fig:PF}
\end{figure}

\noindent \textbf{Pulsation}~~The time evolution of the flux, or equivalently, the pulse structure, could provide us important information about axion conversion. The non-trivial pulse structure of the converted photons from axions arises from the misalignment of the pulsar spin axis with the line of sight (LOS) and the magnetic field's dipole moment. Conventionally, the angle between the pulsar spin axis and the LOS (magnetic dipole moment) is labeled as $\chi$ ($\xi$), as shown in Fig.~\ref{fig:angle}. We also show $\phi$, the pulsar rotation angle. 
In the maximal conversion case, the LOS passes through the pulsar's magnetic equator and $\theta= \pi/2$. Generically, $\theta$ deviates from $\pi/2$, decreasing the converted photon flux. During pulsar's rotation, the direction and intensity of the $B$ field along the LOS is phase-dependent. The converted photon intensity $I$ along the LOS follows
\begin{align}
\label{eq:pulse1}
I&=I_{\rm max}\times \bigg(1-\cos ^2\chi  \cos ^2 \xi   \\\nonumber 
&~~~-\sin ^2 \chi \sin ^2\xi  \cos ^2 \phi +\frac{1}{2} \sin2\chi \sin2\xi \cos
   \phi \bigg)^{\frac{1}{5}} \, ,
\end{align}
where $I_{\rm max}$ is the maximum intensity. The pulse fraction (PF), defined as the ratio of the difference between the maximum and minimum fluxes and the sum of the two, is given by
  \begin{align}
    \label{eq:def-of-pulse-frac}
    \mathrm{PF}     & =
      \begin{cases}
        \frac{\sin^{2/5}(\xi+\chi) - \sin^{2/5}(|\xi-\chi|)}{\sin^{2/5}(\xi+\chi)+\sin^{2/5}(|\xi-\chi|)}, &  \chi + \xi \leq \pi/2 \, , \\
        \frac{1 - \sin^{2/5}(\xi-\chi)}{1+\sin^{2/5}(\xi-\chi)}, &
                                                                   \chi + \xi >\pi/2 \, .
      \end{cases}
  \end{align}
Here PF is independent of axion energy as it arises from magnetic field geometry, which we plot as a function of $\chi$ and $\xi$ in Fig~\ref{fig:PF}. Interestingly, when $\chi+\xi< \pi/2$, there is only one minimum in Eq.~\eqref{eq:pulse1} when $\phi$ goes from 0 to 2$\pi$, corresponding to the single pulse scenario. There are two minima when $\chi+\xi > \pi/2$ since both magnetic poles will approach LOS once per pulsar spin. This will lead to a double pulse structure, which is observed for many pulsars, $e.g.$,~\cite{Kurpas:2024bii,Rigoselli:2019xev,Riley:2019yda}. 

For phase-averaged intensity measurements, the average of photon intensity $\langle I \rangle \equiv \frac{1}{2\pi} \int_0^{2\pi} I d\phi $ introduces an attenuation factor when compared with $I_{\rm max}$, $\left< I \right>/I_{\rm max} < 1$. Numerically, we find the attenuation is mild for most of the parameter space, except where both $\chi$ and $\xi$ are close to zero, as presented in Fig.~\ref{fig:PF}, in which case the conversion is suppressed since $\theta \to 0$. Note that the pulse structure is symmetric under the exchange $\chi\leftrightarrow \xi$. Therefore, there is a degeneracy between the two angles even with very precise pulse measurements. However, the degeneracy will be broken after considering the X-ray polarization discussed below. 
  
\noindent \textbf{Polarization}~~As discussed in Sec.~\ref{sec:pol}, the photon converted from axions will be $\sim100\%$ polarized following the dipole direction. An instantaneous polarization $\sim 1$ in high energy bins could be a strong indicator of the axion origin. However, the magnetic field configuration becomes phase-dependent as long as $\xi\neq 0$. The angle $\Theta$ between the polarization vector and the pulsar spin axis, projected onto the plane of the sky is given as
\begin{equation}
\tan{\Theta} =  \frac{\sin \phi \sin\xi}{\cos \chi \sin \xi \cos\phi-\sin \chi\cos \xi} \, ,
\label{eq:tanTheta}
\end{equation} 
which lifts the degeneracy between $\chi$ and $\xi$.

\begin{figure}
  \includegraphics[width=8.5 cm]{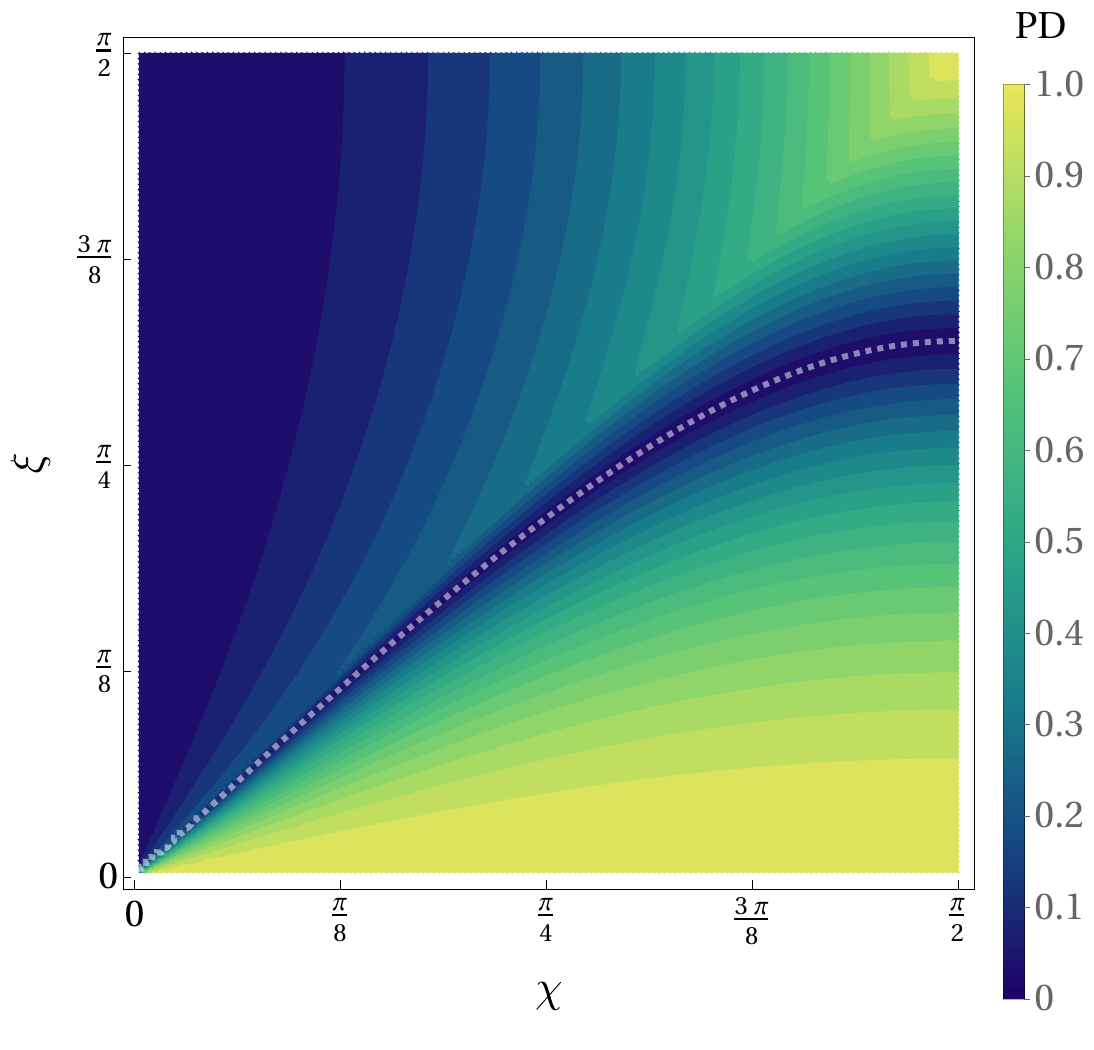}  
\caption{Polarization degree for phase-averaged measurements. The white-dashed curve corresponds to the Stokes parameter $Q=0$, above (below) which we have $Q<0$ ($Q>0$), corresponding to the phase-averaged photons being orthogonal (parallel) to the spin axis, respectively. The color code corresponds to contours of constant polarization degree. 
}
\label{fig:polarization}
\end{figure}

When the measurements have limited time resolution and cannot fully resolve the phase-dependent polarization, the polarization of different phases may cancel each other and attenuate the averaged polarization, similar to the flux. 
Detailed calculations of the phase-averaged polarization angle (PA) and polarization degree (PD) are given in the supplementary material. 
Due to the system's symmetry, the averaged polarization angle will either align with the rotational axis' projection along the LOS or be orthogonal to it, as shown in Fig.~\ref{fig:polarization}.

\section{RX J1856.5-3754: a Case Study}
\label{sec:1856}
The discussions above apply to a wide range of observed NS's. In this section, we will focus on XDINS, a special class of NS, which offers a unique search opportunity for axions in the hard X-ray band. XDINS feature soft thermal X-ray emissions, well characterized by a simple blackbody (BB) spectrum with a temperature between 50 and 100 eV~\cite{Haberl:2003zn, vanKerkwijk:2004qg, Haberl:2004xe,Zane:2005sy}. There are seven of them~\cite{1996Natur.379..233W, Haberl1997, 1998AN....319...97H,Schwope:1998gp, Motch:1999xy, Haberl:1999ck, Zampieri:2001ewa}, nicknamed the ``Magnificent Seven". More recently, several works reported unexpected hard X-ray excesses around and above a keV from XDINS, with enhanced fluxes several orders of magnitude above the BB spectra, based on data from {\it Suzaku}, {\it XMM-Newton}, and {\it Chandra} X-ray telescopes~\cite{DeGrandis:2022kfe, Yoneyama:2017xth, Yoneyama:2018dnh, Dessert:2019dos}. As with all anomalies in the sky, these excesses may be due to pure astrophysical effects such as a non-thermal rotation-powered emission. Yet, it is unclear that an astrophysical explanation could be consistent with all the features of the excesses and other properties of XDINS.

Among the seven of them, RX J1856.6-3754 (J1856 for short later) is the brightest and closest to us, serving as a frequently-observed target for the calibration of the X-ray telescope. Its soft X-ray spectrum is described to a high accuracy by a soft blackbody (SBB) emission with a temperature at infinity $T_{\rm SBB}^\infty \approx 62$ eV~\cite{Burwitz:2001ry, 2003A&A...399.1109B, Sartore_2012}. 
Observations also point to non-trivial pulse profiles of its hard X-ray emission using Neutron Star Interior Composition Explorer (NICER) and {\it XMM-Newton} data. In particular, in the (1.2 -2) keV bin, the PF is 0.29$\pm$0.11 while in the (2-7.5) keV bin, the PF is $\sim 0.5$ but with an even larger error bar due to the systematic error mainly induced by background subtraction and also the lower count rates~\cite{DeGrandis:2022kfe}. In contrast, in the energy bins below 1~keV, the PF is of $\mathcal{O}(0.01)$~\cite{DeGrandis:2022kfe}.

\begin{figure}
\includegraphics[width=7.5 cm]{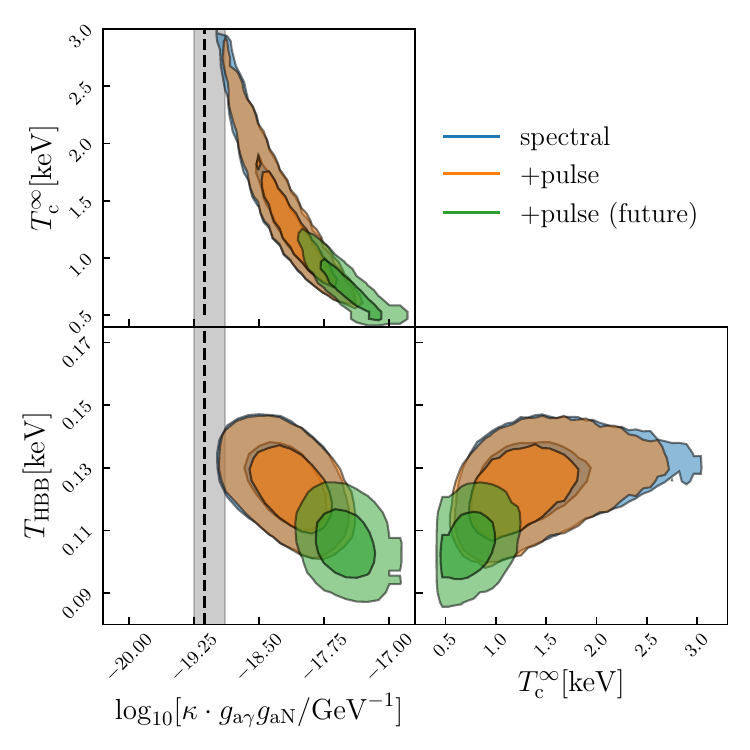}
\caption{Posterior distributions of HBB temperature $T_{\rm HBB}^\infty$, $T_c^{\infty}$, and $g_{a\gamma} g_{aN}$. We absorb the uncertainty due to J1856's mass profile into a dimensionless parameter $\kappa$ and present its impact as a vertical grey band of existing bounds on $\kappa g_{a\gamma} g_{aN}$. More details could be found in App.~\ref{app:estim-uncert-fit}. 
Different contours correspond to the fit with spectrum data only (blue), combined spectrum and pulse data (orange), and combined data with reduced systematic uncertainties in the pulse data (green). The constraint on $g_{a\gamma}g_{aN}$ (vertical-dashed) is from the CAST~\cite{CAST:2024eil} and SN 1987A~\cite{10.1093/ptep/ptaa104} with $\kappa=1$. For $m_a \gtrsim 10^{-5}\,\mathrm{eV}$, the stronger polar cap constraints~\cite{Noordhuis:2022ljw, Caputo:2023cpv} do not directly apply. Gaussian smoothing was applied to the contours with $0.5\,\sigma$ kernel size. 
}
\label{fig:temptemp2D}
\end{figure}

It is suggested in~\cite{Buschmann:2019pfp} that the excess might be accounted by the axion conversion scenario with a light axion and a range of $g_{a \gamma} g_{aN}$ in some tension with existing limits from CAST~\cite{CAST:2024eil} and SN 1987A~\cite{10.1093/ptep/ptaa104,note-3}.
Here we implement a new fit to a larger dataset of J1856 from~\cite{DeGrandis:2022kfe}, including both the flux data above 1.05 keV and the pulsation data in the two high-energy regions (1.2-2.0 and 2.0-7.5 keV), compared to the previous fit with only flux data above 2 keV. 
The X-ray emission has three major contributors~\cite{DeGrandis:2022kfe}: an SBB dominantly determined by the soft spectrum below keV as mentioned above, a hard blackbody (HBB) mostly relevant for the intermediate energy between 1 and 2 keV, and the hard excess $\gtrsim 2$ keV, which we assume to be axion-induced below. Below we simply fix the SBB spectrum following~\cite{DeGrandis:2022kfe}. This choice renders seven free parameters to be fit: $\chi$, $\xi$, rotation phase at a reference time $\phi_0$, the temperature and intensity of the HBB, NS core temperature at infinity $T_c^{\infty}$, and the axion coupling product $g_{a\gamma} g_{aN}$. Details of the fits are listed in the supplementary material. 
Posteriors of three key parameters are shown in Fig.~\ref{fig:temptemp2D}.


From Fig.~\ref{fig:temptemp2D}, we see that the preferred core temperatures from spectrum-only or the combined fits are both around 1.2 keV with large variations. Since the total axion emission rate increases when $T_c^\infty$ increases,
there is an anti-correlation between $T_c^{\infty}$ and $g_{a\gamma} g_{aN}$. The posterior coupling product is only compatible with the current axion bounds when $T_c^\infty\gtrsim 2$~keV. Even higher $T_c^{\infty}$ leads to a stronger tension in the spectrum shape measured. The impact of including current pulse data is not obvious due to its large systematic uncertainties. To better examine the usefulness of the pulse data, we conduct a parallel fit assuming a reduction of the systematic uncertainty to 25\% of the current size. We take a benchmark for the central values to be the current ones.
The result is shown as the green contours in Fig.~\ref{fig:temptemp2D}~\cite{note-4}.
Indeed, this improved pulse precision could potentially exclude the axion explanation of the hard X-ray excess, as it prefers much larger axion couplings.
The power of spectral-pulsation analysis lies in PF's energy dependence. In the large core temperature - small coupling limit, the PF across different high energy bins is expected to become similar in the axion scenario as discussed in the previous section, contradicting what the pulsation data prefers. The tension between the axion coupling constraints and the J1856 data will be greater than $3\sigma$ in this scenario.

\section{Conclusions and Outlook}
\label{sec:con}

In this work, we aim to expand current astrophysical probes of axions with pulsation and polarization measurements on X-rays from pulsars. We focus on the synergy between the three probes of intensity, pulsation, and polarization in disentangling potential axion signals from astrophysical backgrounds. The work is timely given the intriguing hard X-ray excess observed from the XDINS. In particular, we take a critical examination of what information about axions can be extracted from pulsation and polarization measurements. We show that, when fitted against a model of axions coupling to photons and nucleons, the current timing data from \textit{XMM-Newton} is compatible with the theory points preferred by the hard X-ray excess, albeit in some tension with other astrophysical bounds. We demonstrate that a reduction of systematic errors by a factor of 4 in the timing data will reach the precision level needed to rule out the axion explanation of the hard X-ray excess. 

Currently, Imaging X-ray Polarimetry Explorer (IXPE) is in the middle of its 5-year mission to measure the polarization of cosmic X-rays~\cite{2016SPIE.9905E..17W}. So far IXPE has discovered polarized X-rays from magnetars with significantly larger X-ray fluxes~\cite{Taverna_2022, Zane_2023}, and the data has been used to constrain the axion scenario~\cite{Gau:2023rct}. However, it has not provided any information about XDINS. It would be highly interesting if IXPE and future X-ray polarimeters such as~eXTP\cite{eXTP:2018anb}, CATCH~\cite{Li:2022bdm}, and GOSoX~\cite{2021SPIE11822E..0OM}  could collect data from XDINS, {\it i.e.}, time-dependent polarization data, which could shed light on the origin of their mysterious hard X-ray excesses and probe axion physics simultaneously.

Again, we stress that the axion-induced X-ray pulse structures presented are generic for a wide group of sufficiently isolated pulsars. For much brighter sources such as magnetars with both flux and polarization data available, it would be interesting to apply our framework to extract information about axion models from the timing data, which has not been implemented so far and could help resolve astrophysical backgrounds. Alternatively, if an axion is discovered with parameters that can induce observable X-ray signals, this work may serve as a way to probe NS properties. 

\section*{Acknowledgments}

We thank Kfir Blum, Joshua Foster, Fazlollah Hajkarim, Steven Harris, Anirudh Prabhu, Kuver Sinha, Shuxu Yi and Zhen Zhang for useful communication and discussions. We thank Matt Reece for reading the manuscript and providing useful feedback. JF and LL are supported by the NASA grant 80NSSC22K081 and the DOE grant DE-SC-0010010. 
CS acknowledges the receipt of the grant from the Abdus Salam International Center for Theoretical Physics (ICTP), Trieste, Italy.
CS also thanks the hospitality and support of IBS-CTPU and support of CKC for the CTPU-CKC Joint Focus Program: Let there be light (particles) Workshop, where part of this work was performed. 
This work was performed in part at Aspen Center for Physics, which is supported by National Science Foundation grant PHY-2210452.


\bibliography{ref}

\begin{thebibliography}{78}
\bibitem [{ns-()}]{ns-pulsar}%
  \BibitemOpen
  \href@noop {} {}\bibinfo {howpublished} {Below we will use NS's and pulsars
  interchangeably.}\BibitemShut {Stop}%
\end{thebibliography}

\begin{widetext}
  \begin{center}
~~~~
  \end{center}
\end{widetext}
\appendix
\section{Axion-photon Conversion}
\label{app:conversion}
The axion-photon conversion probability is given by, from the leading-order solution to the time-dependent perturbation theory in quantum mechanics~\cite{PhysRevD.37.1237}, 
\begin{equation}
P_{a \to \gamma} (x) = \left| \int_1^{x} d x^\prime \Delta_M(x^\prime) r_0 \, e^{ i \int_1^{x^\prime} d x^{\prime \prime}\left(\Delta_a - \Delta_{\parallel}(x^{\prime\prime})\right) r_0} \right|^2 \, ,
\label{eq:mastereq}
\end{equation}
where 
\beq
\Delta_M &=& \frac{g_{a \gamma} B \sin\theta}{2}, \quad \Delta_a = - \frac{m_a^2}{2 \omega}, \nonumber \\ \quad \Delta_{\parallel} &=& \frac{1}{2} q_{\parallel} \omega \sin^2\theta,~
q_{\parallel} = \frac{7 \alpha}{45 \pi} b^2 \hat{q}_{\parallel}, \nonumber \\  b &=& \frac{B}{B_c}, \quad \hat{q}_{\parallel}= \frac{1+1.2b}{1+1.33b+0.56b^2} \, .
\eeq
In the equations above, $x = r/r_0$ with $r$ the radial distance from the center of the NS and $r_0$ is the NS radius. $g_{a\gamma}$ is the axion-photon coupling, $B$ the NS's external magnetic field, $\theta$ the angle between the direction of axion(photon) propagation and that of the the magnetic field, $m_a$ the axion mass, $\omega$ the photon energy, $\alpha=1/137$ the electromagnetic fine structure constant, and $B_c = m_e^2/e= 4.414 \times 10^{13}$ G with $e$ the electric charge is the critical magnetic field strength. The $q$ factor takes into account of the vacuum polarization effects and the general fitting formula above is taken from~\cite{Potekhin_2004,Lai:2006af}. 

For our purpose, it is sufficient to consider the dipole component of the NS magnetic field. At the magnetic equator, the field intensity is approximated as:
\beq
|B(r)| = B_0 x^{-3} \, ,
\eeq
with $B_0$ being the dipole field strength at the NS surface. 

While it is possible to evaluate the integration fully numerically, we will derive an approximate analytic result which agrees very well with the pure numerical result in the parameter space of interest.  
The integration inside the exponential in Eq.~\eqref{eq:mastereq} could be approximated as 
\beq
& &\int_1^{x^\prime} d x^{\prime \prime}\left(\Delta_a - \Delta_{\parallel}(x^{\prime\prime})\right) r_0 \nonumber \\ &=&  \int_1^{x^\prime} d x^{\prime \prime}\Delta_a  r_0\left( 1 + \left(\frac{x_c}{x^{\prime\prime}}\right)^6 \hat{q}_\parallel (x^{\prime\prime})\right)  \, , \\
&\approx& \Delta_a  r_0 \left(x^\prime - \frac{x_c}{5} \left(\frac{x_c}{x^{\prime}}\right)^5 \right) \, ,
\eeq
where
\beq
x_{c} = \left(\frac{7 \alpha}{45 \pi}\right)^{1/6} \left(\frac{B_0 \omega |\sin \theta|}{B_c m_a}\right)^{1/3} \, .
\eeq
In the second step, we make the approximation that $\hat{q}_\parallel \approx 1$, which facilitates the integration analytically. The approximation works for small $B_0\lesssim B_c$ cases very well ($e.g.$, for when $|B|=1\times 10^{13}$~G, $\hat{q}_\parallel=1.2$). The approximation could also be extended for NS with very large fields such as the magnetars~\cite{Fortin:2018aom} as the conversion mainly happens at the large $x$ region with small field strength as shown in Fig.~1 of the main text.  We also ignore the spatial independent contributions from the integration end $x^{\prime\prime} = 1$ since they only contribute phase terms and will drop off in the final probability. Plugging the approximated exponent above into Eq.~\eqref{eq:mastereq}, we get
\beq
P_{a \to \gamma}  &\approx & \left| \int_1^{x} d x^\prime \Delta_M(x^\prime) r_0 \, e^ {i\Delta_a  r_0 \left(x^\prime - \frac{x_c}{5} \left(\frac{x_{c}}{x^{\prime}}\right)^5 \right)} \right|^2  \, \nonumber \\
& = & \frac{(\Delta_M(1)r_0)^2}{x^4_c} \left| \int^{x/x_c}_{1/x_c}  \frac{d t}{t^{3}} \, e^{i \Delta_a r_0 x_c \left(t-\frac{1}{5t^5}\right) }\right|^2 \nonumber \\
&\approx&  \frac{(\Delta_M(1)r_0)^2}{x^4_c} \left| \int^{x/x_c}_{1/x_c}  \frac{d t}{t^{3}} \, e^{i \Delta_a r_0 x_c \left(-\frac{1}{5t^5}\right) }\right|^2  \nonumber \\
&=&  \frac{(\Delta_M(1)r_0)^2}{5^{\frac{6}{5}} |\Delta_a r_0 x_c|^{\frac{4}{5}}x^4_c } \times \nonumber \\ & & \left |\Gamma \left(\frac{2}{5}, \frac{i \Delta_a r_0 x_c}{5 (x/x_c)^5} \right)-\Gamma \left(\frac{2}{5}, \frac{i \Delta_a r_0 x_c}{5 (1/x_c)^5} \right) \right |^2
\label{eq:secondapprox}
\eeq
where in the second line, we change the variable from $x^\prime$ to $t= x^\prime/x_c$, while in the third line, we make an approximation by omitting the $\Delta_a r_0 x^\prime =\Delta_a r_0 x_c t $ term in the exponent and only keep the fast oscillating $1/t^5$ term. The asymptotic conversion probability is then given as~\cite{Fortin:2018aom}
\beq
P_{a \to \gamma} (\infty)&\approx&(\Delta_M(1)r_0)^2 \frac{1}{x^4_c} \left| \int^{\infty}_{0} d t \frac{1}{t^{3}} \, e^{i \Delta_a r_0 x_c \left(-\frac{1}{5t^5}\right) } \right|^2 \nonumber \\ 
&=& (\Delta_M(1)r_0)^2 \frac{1}{x^4_c} \frac{\Gamma(2/5)^2}{5^{6/5} |\Delta_a r_0 x_c|^{4/5} } \, , \label{eq:asymptotic}
\eeq
where a further approximation is made by ignoring the negligible contribution from the integration range $(0, 1/x_c)$. The conversion radius $r_{\rm con}$, which is introduced in Eq.~II.1 in the main text is then defined to be the radius when $P_{a\to \gamma}(x_{\rm con})=P_{a\to \gamma}(\infty)/2$.

\section{Geometry of NS Observables}
\label{app:geometry}
In this section, we show definitions of 
quantities for the pulsation and polarization measurements. We also present derivations of the formulas in Sec.~III of the main text, which describes how the axion-converted X-ray signal modulates with the phase $\phi$.

\subsection{Pulse Structure}
\label{sec:pulse-structure}
Because of the NS rotation and the misalignment between the spin axis, the LOS, and the magnetic axis, the conversion probability exhibits periodic oscillations. Given an instantaneous configuration of $\xi, \chi$, and $\phi$, the corresponding $P_{a\to \gamma}$ only depends on the combination $(B_0 \sin\theta)^{2/5}$ as shown in Eq.~\eqref{eq:asymptotic}. In the dipole field approximation, the surface field $B_0$ and the angle $\theta$ between the propagation direction and the local magnetic field only depends on the angle $\beta$ between the LOS and the dipole axis. It is straightforward to obtain $B_0 \sin \theta \propto \sin\beta$.

\begin{figure}[h!]
\includegraphics[width=7cm]{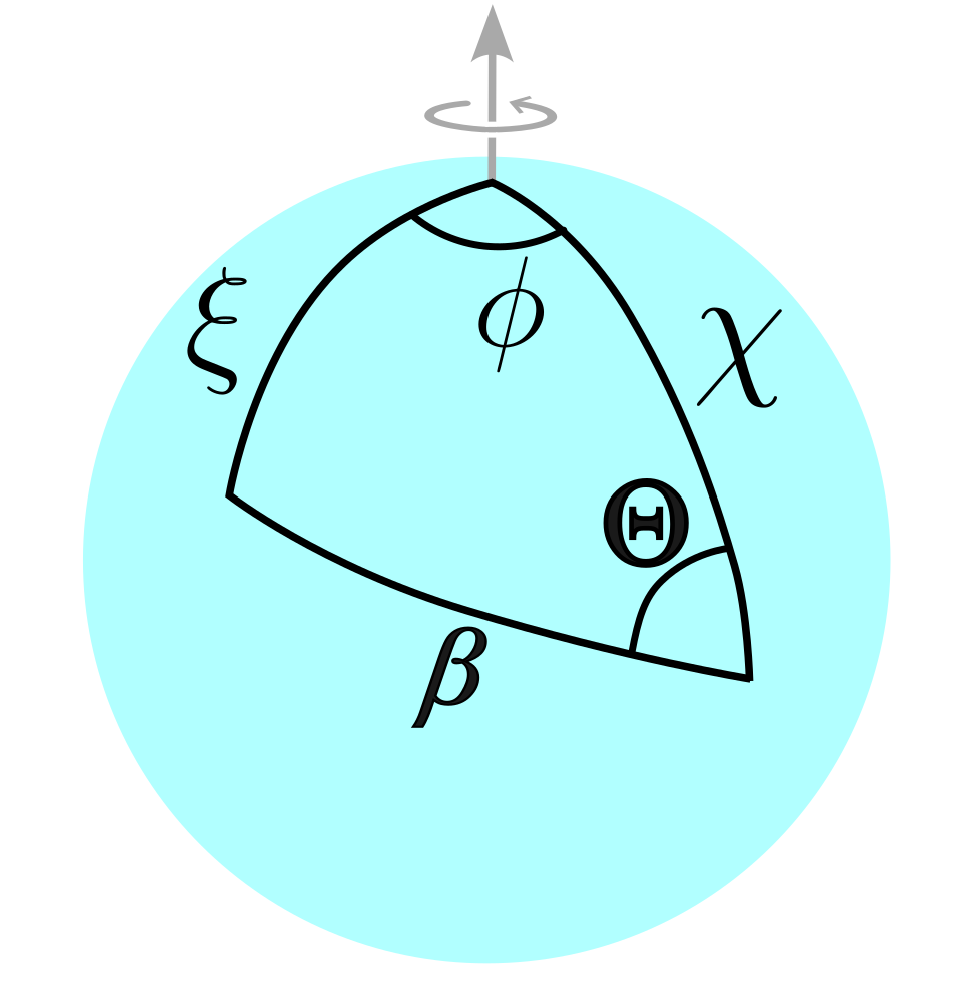}
\caption{\label{fig:geometry1} {The spherical triangle for an instantaneous configuration. All the angles $\xi, \chi, \beta$ defined based on the spherical center corresponds to an arc on the surface, thus a side of the triangle.}}
\end{figure}

In Fig.~\ref{fig:geometry1}, the relevant quantities appear in the same spherical triangle. According to the spherical law of cosine, we have
\begin{equation}
\cos \beta = \cos \xi \cos \chi + \sin \xi \sin \chi \cos\phi~.
\end{equation}
The pulsation is then a result of the following dependence of the axion-to-photon conversion probability, $P_{a \rightarrow \gamma}$ on the rotation phase $\phi$, 
\begin{align}
  P_{a\rightarrow \gamma}
  & \propto (B_0\sin\theta)^{2/5} \cr
  & \propto \sin\beta^{2/5}
    \cr
  & = 
    [1-(\cos\xi \cos\chi +\sin\xi \sin\chi\cos\phi)^2]^{\frac{1}{5}} \, . 
    \label{eq:intensity1}
\end{align}
The local extremum could be found by setting $dP_{a\rightarrow \gamma} /d \cos \phi=0$, 
\begin{align}
  \frac{d P_{a\rightarrow \gamma}}{d \cos \phi}
      = \frac{1}{5}\, \frac{P_{a\rightarrow \gamma} }{f(\xi,\chi, \phi)} (-2\sin^2 \xi \sin^2 \chi)(\cot \xi \cot \chi + \cos \phi) \, .
\end{align}
Since $P_{a\rightarrow \gamma}$ and $f(\xi,\chi,\phi)$ are both positive definite, the actual sign of the derivative is determined by comparing $\cos\phi$ with $\cot \xi \cot \chi$.

\noindent \textbf{Small misalignment}~~When $\cot \xi \cot \chi > 1$, or equivalently $\xi+\chi<\pi/2$,  the derivative is monotonic with respect to $\cos \phi$. Therefore, the maximum is achieved at $\phi=\pi$ and the minimum at $\phi=0$ in a $[0, 2\pi)$ range. Using the conversion rate given in Eq.~\eqref{eq:intensity1}, we get the PF as
\begin{align}
  \mathrm{PF}
  & = \frac{\sin^{2/5}(\xi+\chi) - \sin^{2/5}(\xi-\chi)}{\sin^{2/5}(\xi+\chi)+\sin^{2/5}(\xi-\chi)} \, .
\end{align}

\noindent \textbf{Large misalignment}~~When $0 \leq \cot \xi \cot \chi \leq 1$, or equivalently $\xi+\chi>\pi/2$, the extremum is achieved at $\cos \phi = - \cot \xi \cot \chi$. This corresponds to $f(\xi,\chi,\phi)=1$, a global maximum. In this regime, there are two local minima at $0$ and $\pi$ in the $\phi \in [0, 2\pi)$ range, with the former being the global minimum. 
Using the global maximum and minimum, the PF is 
\begin{align}
  \mathrm{PF}
  & = \frac{1 - \sin^{2/5}(\xi-\chi)}{1+\sin^{2/5}(\xi-\chi)} \, .
\end{align}

\subsection{Polarization}
\label{ssec:geometry}
\noindent \textbf{Instantaneous Photon Vector}~~We first show the coordinate setup at a given moment and the orientation of the photon's polarization vector, which will be related to the time-dependent quantities such as the PA. We set up the coordinate such that we have the following unit vectors:
\begin{align}
  \label{eq:coord-setup}
  \hat n_s
  & = (0, 0, 1) \, ,
  \\
  \hat n_B
  & = (\sin \xi \cos \phi, \sin \xi \sin \phi, \cos \xi) \, ,
  \\
  \hat n_k
  & = (\sin \xi, 0, \cos \xi) \, ,
\end{align}
where $\hat n_s, \hat n_B, \hat n_k$ are along the spin, magnetic field, and LOS, respectively. Following the convention~\cite{zotero-24746,Poutanen:2020lom,Taverna:2022jgl}, we set up the polarization basis of the plane perpendicular to the LOS as follows:
\begin{align}
  \hat e_1
  & = \frac{\hat n_s - (\hat n_s\cdot \hat k) \hat k}{|\hat n_s - (\hat n_s\cdot \hat k) \hat k|} = (-\cos \xi, 0, \sin \xi) \, ,
  \\
  \hat e_2
  & = \frac{\hat n_k \times \hat n_s}{|\hat n_k \times \hat n_s|} = (0, -1, 0) \, .
\end{align}
The axion-induced photons have a polarization vector given by
\begin{align}
  \cos \Theta 
  & = \hat n_B \cdot \hat e_1 \, ,
    \\
  \sin \Theta 
  & = \hat n_B \cdot \hat e_2 \, ,    
\end{align}
which leads to Eq.~III.3 of the main text about $\tan \Theta$. 
\noindent \textbf{Polarization Variables}~~Having set up the geometry for the polarization vector of the photon, we can cast it into the Stokes parameters. For photons with a potential vector $\mathbf{A}$, as we do not expect any phase shift between $A_1$ and $A_2$, there is no circularly polarized mode, \textit{i.e.} $V=0$. In the basis of $\hat e_1$ and $\hat e_2$, the Stokes parameters are given by
\begin{align}
  I & = \left <A_1^2 \right >  + \left <A_2^2 \right > = \left < |\mathbf{A}|^2 \right > \, ,
  \cr
  Q & = \left <A_1^2 \right >  - \left <A_2^2 \right > =  \left <|\mathbf{A}|^2\cos (2 \Theta) \right > \, ,
      \cr
      U & = \left <A_\alpha^2 \right >  - \left <A_\beta^2 \right >
          \cr
  & = \left <|\mathbf{A}|^2 \cos^2(\Theta-45^\circ) \right >  -
          \left <|\mathbf{A}|^2 \sin^2(\Theta-45^\circ) \right >
          \cr
          & = \left <|\mathbf{A}|^2 \sin (2 \Theta) \right > \, , 
\end{align}
where $A_\alpha$ and $A_\beta$ are in the basis of $\{\hat e_\alpha, \hat e_\beta\}$ that is $45^\circ$ rotation over $\{\hat e_1, \hat e_2\}$.
By definition, the PA is given by
\begin{align}
  \label{eq:PA-time-avg}
  \tan(2\mathrm{PA})
  & = \frac{U}{Q}
    = \frac{\left < |\mathbf{A}|^2 \sin(2\Theta) \right >}
    {\left <|\mathbf{A}|^2 \cos(2\Theta) \right >} \, .
\end{align}
For a measurement that spans small enough in the NS phase, $\left < |\mathbf{A}|^2\cos(2\Theta ) \right > \approx |\mathbf{A}|^2 \cos(2\Theta)$, and $\left < |\mathbf{A}|^2\sin(2\Theta) \right > \approx |\mathbf{A}|^2 \sin(2\Theta)$, hence, we have $\mathrm{PA}=\Theta$. For measurements averaged over the NS rotation phase, the two do not coincide in general. 

%
The PD is defined as
\begin{align}
  \mathrm{PD}
  \label{eq:PD-time-avg}
  & = \frac{\sqrt{Q^2 + U^2}}{I}
  \\
  & = \bigg [\left <|\mathbf{A}|^2 \cos(2\Theta) \right >^2 + \left <|\mathbf{A}|^2 \sin(2\Theta) \right >^2  \bigg ]^{1/2}/ \left < |\mathbf{A}|^2 \right > . \notag
\end{align}
Again, in the instantaneous limit, it is obvious that PD=1 due to the fact that axion-induced photons are all polarized along the magnetic field direction.

In the case of a dipole magnetic field, the intensity has a phase dependence given by
\begin{align}
|\mathbf{A}|^2& \propto \bigg(1-\cos ^2\chi  \cos ^2 \xi   \\\nonumber 
&~~~-\sin ^2 \chi \sin ^2\xi  \cos ^2 \phi +\frac{1}{2} \sin2\chi \sin2\xi \cos
   \phi \bigg)^{1/5}~.  
\end{align}
One can use this to carry out the phase average in Eqs.~(\ref{eq:PA-time-avg})(\ref{eq:PD-time-avg}).

\section{Details of the Numerical Fit}
\label{app:fitdetails}
In our framework, the total intensity of photons converted from axions is proportional to $|g_{a\gamma}g_{aN}|^{2}$.
Since axion production is related to many NS and nuclear physics parameters that are not precisely known, in this work, we only aim to provide a benchmark value of axion production. In particular, we follow the calculation in~\cite{Fortin:2021sst,Gau:2023rct} with the single pion exchange approximation.\endnote{For computations based on alternative approximations, see~\cite{Hardy:2024gwy}.} We also assume that there is no significant Cooper-pair breaking contribution, similar to~\cite{Buschmann:2019pfp}, and the dominant contribution comes from the $n+n\to n+n+a$ bremsstrahlung process. 

\subsection{Inputs}
\label{sec:inputs}

We use the following flat priors for the Markov chain Monte Carlo (MCMC) runs.
\begin{align}
  \label{eq:priors-model}
  \xi
  & \in (0, \pi/2) \, , \cr
    \chi
  & \in (0, \pi/2) \, ,
    \cr
    \phi_0
  & \in (0, \pi) \, , \cr
    T_{\rm HBB}^\infty ~[\mathrm{keV}] 
  & \in (0.04, 1.0) \, ,
    \cr
    R_{\rm HBB} ~[\mathrm{km}] 
  & \in (0.01, 12) \, ,
    \cr
    T_{c}^{\infty}~[\mathrm{keV}]
  & \in (0.1, 3) \, ,
    \cr
    \log_{10}(g_{aN}g_{a\gamma} /\mathrm{GeV}^{-1})
  & \in (-22, -17)  \, ,~~~  
\end{align}
where $R_{\rm HBB}$ parametrizes the attenuation in the X-ray emitted by HBB as $(R_{\rm HBB}/r_0)^2$ with $r_0$ the NS radius. As such, $R_{\rm HBB} = r_0$ corresponds to no reduction.

We fix the other parameters to the following benchmark values. See Sec.~\ref{app:estim-uncert-fit} of the supplementary material for more discussions on the uncertainties related to these parameters.
\begin{align}
\label{eq:benchmark}
  r_0
  & = 12\,\mathrm{km} & \text{\cite{Burgio:2021vgk}} \, , \cr
          M_{\rm NS} 
 & = 1.4\, M_\odot  & \text{\cite{Ho:2006uk}} \, ,   \cr
                      D_{\rm NS}
  & = 123\, \mathrm{pc} & \text{\cite{Walter:2010ht}} \, , \cr
                        B_0
   & = 1.5 \times 10^{13}\,\mathrm{G} & \text{\cite{vanKerkwijk:2007jp}} \, , \cr
    T_{\rm SBB}^\infty 
 & = 62\,\mathrm{eV},  & \text{\cite{DeGrandis:2022kfe}} \, , \cr
                         R_{\rm SBB}
  & = 12\, \mathrm{km}  & \text{\cite{DeGrandis:2022kfe}}  \, ,
\end{align}
where $r_0$ is the radius of J1856, $M_{\rm NS}$ is its mass, $D_{\rm NS}$ is the distance to J1856, $T_{\rm SBB}^\infty$ is the temperature of its SBB component, and $R_{\rm SBB}$ is the effective radius of the SBB component. 
We assume the dipole profile for the magnetic field. $B_0$ is the surface magnitude of the magnetic dipole at the equator. 
We anchor the temperature of the SBB component using the posterior of \cite{DeGrandis:2022kfe} since it is constrained mostly by spectral data below 1~keV. 
In the MCMC, we assume a homogeneous density profile with the mass density $\rho_{\rm NS} = M_{\rm NS}/(\frac{4 \pi}{3} r_0^3)$ to compute the Fermi momentum of the neutrons. This underestimates the axion flux compared to the result taking into account the more accurate mass profile and NS's equation of state (EoS). We correct this by computing the benchmark using the package \texttt{CompactObject}~\cite{Huang:2024rfg}. More explicitly, the axion luminosity from neutron bremsstrahlung, $L_a$, is proportional to the following integral~\cite{Gau:2023rct}
\begin{align}
  \label{eq:La-grav}
  L_a \propto
  &
    \int_0^{r_0} \,
    \frac{dr \, r^2}{\left(1-\frac{2G M_{\rm NS}(r)}{r} \right )^{1/2}} \, p_{F}\, F \left ( \frac{m_\pi}{2p_F} \right ) \mathrm{e}^{-4\phi(r)},
\end{align}
where $F(x) = 4 - (1+x^2)^{-1} - 5\,x\, \mathrm{tan}^{-1}\left  (x^{-1}\right )
+ 2x^2(1+2x^2)^{-1/2}\, \mathrm{tan}^{-1}\left [ (1+2x^2)^{-1/2} \right ]$, $\phi(r)$ is the gravitational potential, $p_F$ the neutron's Fermi momentum, $G$ the Newton constant, and $m_\pi$ the pion mass.
By solving the EoS explicitly for J1856, we see a factor of 2.98 enhancement in the axion luminosity compared to the homogeneous density profile. The uncertainty in the total mass of J1856 leads to a variation of the axion luminosity, which impact we will quantify in Sec.~\ref{app:estim-uncert-fit} of the supplementary material.


\subsection{Data}
\label{sec:data}

The X-ray spectrum and pulse data are adopted from~\cite{DeGrandis:2022kfe}. More specifically, the spectrum data is from the observation of {\it XMM-Newton} EPIC-pn camera during 2002 to 2022 
operating under the Prime Small Window mode. Some light selection criteria were imposed in the analysis \cite{DeGrandis:2022kfe}.

The timing data is a combination of the data from the EPIC-pn camera and that from the NICER mission acquired during a 7-day high cadence long exposure in 2019. However, since the 60~ks NICER data only spans the energy range of 0.2-1~keV, we only use the pulsation data from EPIC-pn observations. This also avoids the complication of combining different observations in producing the energy-resolved pulsation data.

Since the phase-average intensities of the pulse data are related to the spectrum, there could be potential double counting when combining the spectrum and pulse data. To avoid double-counting, we introduce two nuisance parameters for the normalization of the pulsation data in the two energy bins we use in our fit. We will show this more explicitly in Sec.~\ref{sec:likelihood} of this supplementary material.

\subsection{Likelihood}
\label{sec:likelihood}

In dealing with the spectral and pulsation data, we assume a diagonal covariance matrix, \textit{i.e.} no correlation between energy bins.
The log-likelihood for the spectral data is given by
\begin{align}
  -2\,\mathrm{Log}(\mathrm{likelihood})
  & =
    \sum_i
    \left [ \frac{\mu^{\rm th}(E_i) - \mu^{\rm exp}(E_i)}{\sigma(E_i)}  \right ]^2 \, ,
\end{align}
where $\mu^{\rm th}$ ($\mu^{\rm exp}$) is the theoretical prediction (observation) of the spectral number intensity in the unit of $\mathrm{s}^{-1}\,\mathrm{cm}^{-2}\,\mathrm{keV}$, and $\sigma(E_i)$ is the reported uncertainty. The sum is over the energy bins from 1.05~\textrm{keV} to 7.50~\textrm{keV}.

The log-likelihood for the pulsation data is given by
\begin{align}
  \label{eq:log-likelihood}
  & -2 \, \mathrm{Log}(\mathrm{likelihood})
  \\
  & =
    \sum_{i=1}^{2}
    \mathrm{min}_{\xi_i}
    \left [
    \sum_j
    \left ( \frac{\xi_i\lambda_{i,j}^{\rm th} - \lambda_{i,j}^{\rm exp}}{\sigma_{i,j}}  \right )^2
    \right ],
    \notag
\end{align}
where index $i$ goes over the two energy bands $\Delta E_i$ with $\Delta E_1 = (1.2, 2.0) \,\mathrm{keV}, \Delta E_2 = (2.0,7.5) \,\mathrm{keV}$, and index $j$ goes through the rotation phase of J1856.  $\lambda^{\rm th} (\lambda^{\rm exp})$ is the theoretical prediction (observation) of the photon counting rate in the energy band, and  $\sigma$ the standard deviation in each time interval. The parameter $\xi_i$ accounts for the normalization of the related energy bin such that the spectral shape is not explicitly used here. This avoids double-counting the constraining power of the spectral data in the fit.

The theoretical predictions of both the spectral number intensity and the photon counting rate include three components
\begin{itemize}
\item the soft blackbody contribution with $T_{\rm SBB}^\infty$ and the effective radius $R_{\rm SBB}$, 
\item the hard blackbody contribution with $T_{\rm HBB}^\infty$ and $R_{\rm HBB}$, 
\item the axion-induced component with spectrum determined by the core temperature $T_{c}^{\infty}$ and normalization by the axion coupling $g_{aN} g_{a\gamma}$.
\end{itemize}
More explicitly, the spectral prediction is given by
$ \mu^{\rm th}_i = \mu^{\rm soft}_i + \mu^{\rm hard}_i + \mu^{\rm axion}_i$, where
\begin{widetext}
\begin{align}
  \mu^{\rm soft}_i
  & = 
    \frac{1}{E_{i+1} - E_{i}}
    \int_{E_i}^{E_{i+1}} dE \, \int_0^{\Omega_{\rm NS}}d\Omega \,
    \frac{I_{\rm soft}(E, T_{\rm SBB}^\infty)}{E} \, \left ( \frac{R_{\rm SBB}}{r_0} \right )^2,
  \\
  \mu^{\rm hard}_i
  & = 
    \frac{1}{E_{i+1} - E_{i}}
    \int_{E_i}^{E_{i+1}} dE \, \int_0^{\Omega_{\rm NS}} d \Omega \,
    \frac{I_{\rm hard}(E, T_{\rm HBB}^\infty)}{E} \, \left ( \frac{R_{\rm HBB}}{r_0} \right )^2,
  \\
  \mu^{\rm axion}_i
  & =
    \int_0^{2\pi} \frac{d\phi}{2\pi}
    \frac{1}{E_{i+1} - E_{i}}
    \int_{E_i}^{E_{i+1}} dE \,
    \int_0^{\Omega_{\rm NS}} d \Omega \,
    \frac{    I_{\rm axion}(E, T_c^\infty, g_{aN})}{E}\cdot p_{a\rightarrow \gamma}( E, g_{a\gamma}, B_0, r_0, \xi, \chi, \phi),    
\end{align}
\end{widetext}
where the integration over $d\Omega$ covers the solid angle spanned by J1856 seen by the observer on Earth, $I_{\rm soft, hard}$ is the specific intensity of a blackbody with temperature $T_{\rm SBB}^\infty (T_{\rm HBB}^\infty)$, respectively. The specific intensity of the X-ray signal is related to the produced axion intensity and the axion-to-photon conversion probability. Without considering the gravitational potential, the specific intensity of axions is related to the spectral emissivity by $I_{\rm axion} = \left ( \frac{r_0}{3 \pi} \right ) \,\frac{d\epsilon_{aN}}{dE} (g_{aN}, T_c^\infty, \rho_{\rm NS}) $, which is well computed in the literature. See, for example, Eq.~(4.1) in \cite{zotero-24826}. The gravitational effect is then accounted for using the integration in Eq.~(\ref{eq:La-grav}). 

Similarly, the energy-averaged time-resolved counting rate is given by $ \lambda^{\rm th}_{i,j} = \lambda^{\rm soft}_{i,j} + \lambda^{\rm hard}_{i,j} + \lambda^{\rm axion}_{i,j}$
\begin{widetext}
  \begin{align}
    \lambda^{\rm soft}_{i,j}
    & =
      \frac{1}{\Delta E_i} \int_{E_i}^{E_i + \Delta E_i}dE\,
      \frac{1}{\Delta \phi_j} \int_{\phi_j}^{\phi_j+\Delta \phi_j}d\phi\,
      \int_0^{\Omega_{\rm NS}} d\Omega\, \frac{I_{\rm soft}}{E}
      \left ( \frac{R_{\rm SBB}}{r_0} \right )^2 
      A(E) \, q(E) \, ,
    \\
    \lambda^{\rm hard}_{i,j}
    & =
      \frac{1}{\Delta E_i} \int_{E_i}^{E_i + \Delta E_i}dE\,
      \frac{1}{\Delta \phi_j} \int_{\phi_j}^{\phi_j+\Delta \phi_j}d\phi\,
      \int_0^{\Omega_{\rm NS}} d\Omega\, \frac{I_{\rm hard}}{E}
      \left ( \frac{R_{\rm HBB}}{r_0} \right )^2 
      A(E) \, q(E) \, ,
    \\
    \lambda^{\rm axion}_{i,j}
    & =
      \frac{1}{\Delta E_i} \int_{E_i}^{E_i + \Delta E_i}dE\,
      \frac{1}{\Delta \phi_j} \int_{\phi_j}^{\phi_j+\Delta \phi_j}d\phi\,
      \int_0^{\Omega_{\rm NS}} d\Omega\, \frac{I_{\rm axion}}{E}
      p_{a\rightarrow \gamma}
      A(E) \, q(E)  \,  ,
  \end{align}
\end{widetext}
where $A(E)$ is the effective area of the detector and $q(E)$ the efficiency. The profile of EPIC-pn is taken from the EPIC calibration status \texttt{XMM-SOC-CAL-TN-0018} and the mirror calibration release \texttt{CAL-SRN-0205-1-0}. 


\subsection{Posterior}
\label{sec:posterior}
We make three separate runs using different datasets. We first run with spectral data only, $\mathcal{Q}_1$, and spectral data plus the pulsation data $\mathcal{Q}_2$. To estimate the amount of improvement in systematic error needed for the pulsation data to become more constraining than the spectrum data, we have a third run reducing the error budget to 25\% of the current measurement, assuming that the central values are unchanged. We denote this run as $\mathcal{Q}_3$.

We run MCMC using \texttt{emcee}~\cite{Foreman-Mackey:2012any} with 100 walkers, each with 50000 steps, so effectively, the chain length is $5\times 10^6$. We set the convergence condition to be the final chain size reaching at least 50 times the autocorrelation length.
We show the 1D posterior in Tab.~\ref{tab:1D-posterior} after marginalizing over the rest of the parameters. For $\mathcal{Q}_1$, the data has no preference in either $\chi$ or $\xi$ as the pulse information is not included. Both angles are uniformly distributed between 0 to 2$\pi$. From $\mathcal{Q}_1$ to $\mathcal{Q}_2$, the uncertainties of $\chi$ and $\xi$ are only reduced mildly due to the large systematic uncertainties of the pulse data. The uncertainties further reduce in scenario $\mathcal{Q}_3$. We stress that the results of $\mathcal{Q}_3$ depend on the new central values after the improvement of the resolution, which we assume to be the same as the current ones. With the limited data resolution and moderate parameter space dimensionality, the posterior we obtain is smoothly distributed and shows satisfactory convergence. For more precise data and more papameters to be determined, due to the non-trivial polarization-pulse pattern of the axion-induced signal, the posterior distribution may become complicated and difficult to sample. Advanced techniques such as nested sampling (see $e.g.$~\cite{Buchner:2021kpm}) will be needed.

\begin{table}[h]
  \centering
  \begin{tabular}{c|c c c c c }
    \hline 
    & $\mathrm{Log}(g\cdot g)$ & $T_c^\infty$ [keV] & $T_{\rm HBB}^\infty$ [keV] & $\xi$ & $\chi$ \\
    \hline
    $\mathcal{Q}_1$ &  $-18.25^{+0.29}_{-0.33}$ & $1.30^{+0.52}_{-0.32}$  & $0.12^{+0.01}_{-0.01}$  & $0.79^{+0.52}_{-0.53}$& $0.79^{+0.54}_{-0.53}$ \\
    $\mathcal{Q}_2$ &  $-18.19^{+0.28}_{-0.33}$ & $1.26^{+0.50}_{-0.31}$  &  $0.12^{+0.01}_{-0.01}$& $0.72^{+0.46}_{-0.41}$& $0.73^{+0.46}_{-0.42}$ \\
    $\mathcal{Q}_3$ &  $-17.59^{+0.22}_{-0.22}$ & $0.76^{+0.19}_{-0.15}$  & $0.11^{+0.01}_{-0.01}$  & $0.42^{+0.24}_{-0.20}$& $0.41^{+0.23}_{-0.20}$ \\
    \hline 
  \end{tabular}
  \caption{\label{tab:1D-posterior}
    The mean and $1\sigma$ values of the model parameters, fitted to various data combinations, $\mathcal{Q}_1, \mathcal{Q}_2$ and $\mathcal{Q}_3$.
    The column of $\mathrm{Log}(g\cdot g)$ is for $\mathrm{Log}_{10}(g_{a\gamma} g_{aN}/\mathrm{GeV}^{-1})$. As is discussed in the main text, the shift of the central values in the posterior of $\mathcal{Q}_3$ is purely due to the assumption of the central values of the measurement, as an indicator that the pulsation data becomes more constraining than the spectrum data. 
  }
\end{table}

\begin{figure}[t]
\includegraphics[width=7.5 cm]{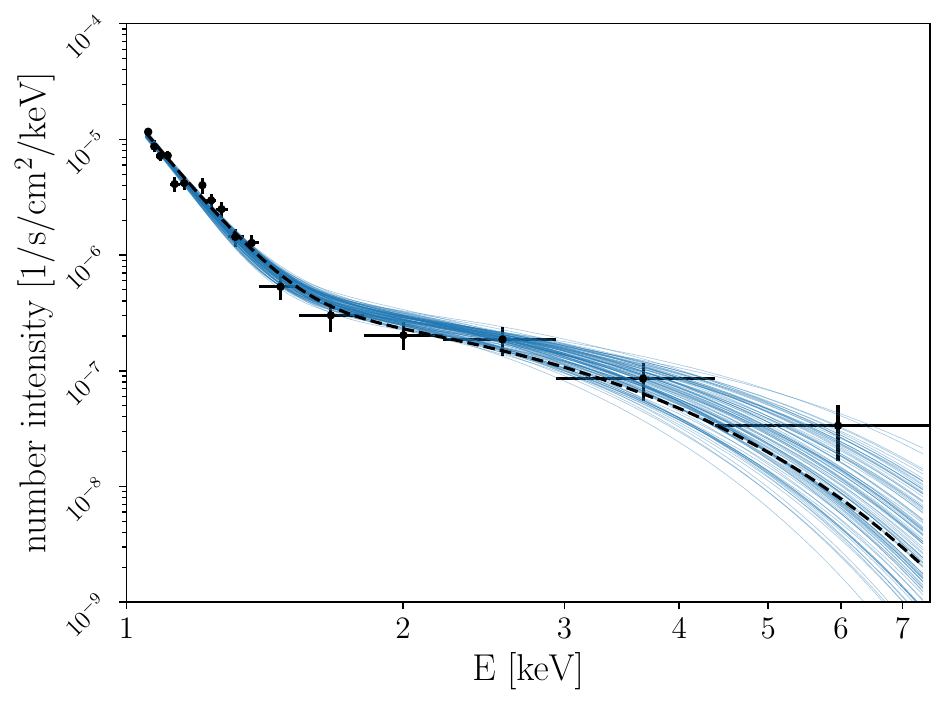}
\includegraphics[width=7.5 cm]{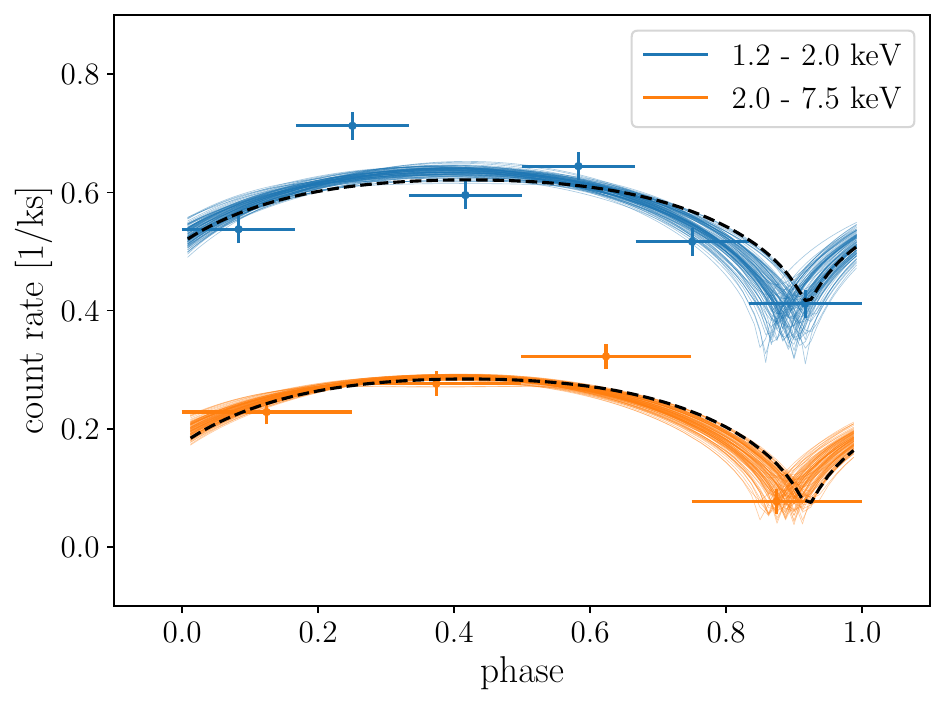}
\caption{100 theory points within $1~\sigma$ of the best fit point, randomly picked. 
The large scattering of the pulsation mock data is due to the assumption that the central values do not shift from the current values.}
\label{fig:bestfit1}
\end{figure}

\subsection{Best Fit Point}
\label{sec:best-fit-point}
It is observed that in Run~$\mathcal{Q}_3$, the reduction of the systematic error to 25\% marks the constraining power of the folded pulsation time-series dominating over the spectrum data. We visualize this by randomly picking out 100 theory points that are within $1~\sigma$ of the best fit point. We show this in Fig.~\ref{fig:bestfit1}. It is observed that with this benchmark assumption of the central values for the pulsation data, it pulls toward the direction of small core temperature - large axion coupling, in order to generate a larger PF for the 2.0 - 7.5 keV band compared to the 1.2 - 2.0 keV band. This is achieved with a penalty of worsening the fit to the spectral data in the high energy bins above 3 keV.

\section{Estimate of the Uncertainty of the Fit}
\label{app:estim-uncert-fit}
\begin{figure}[t]
  \centering
  \includegraphics[width=.45\textwidth]{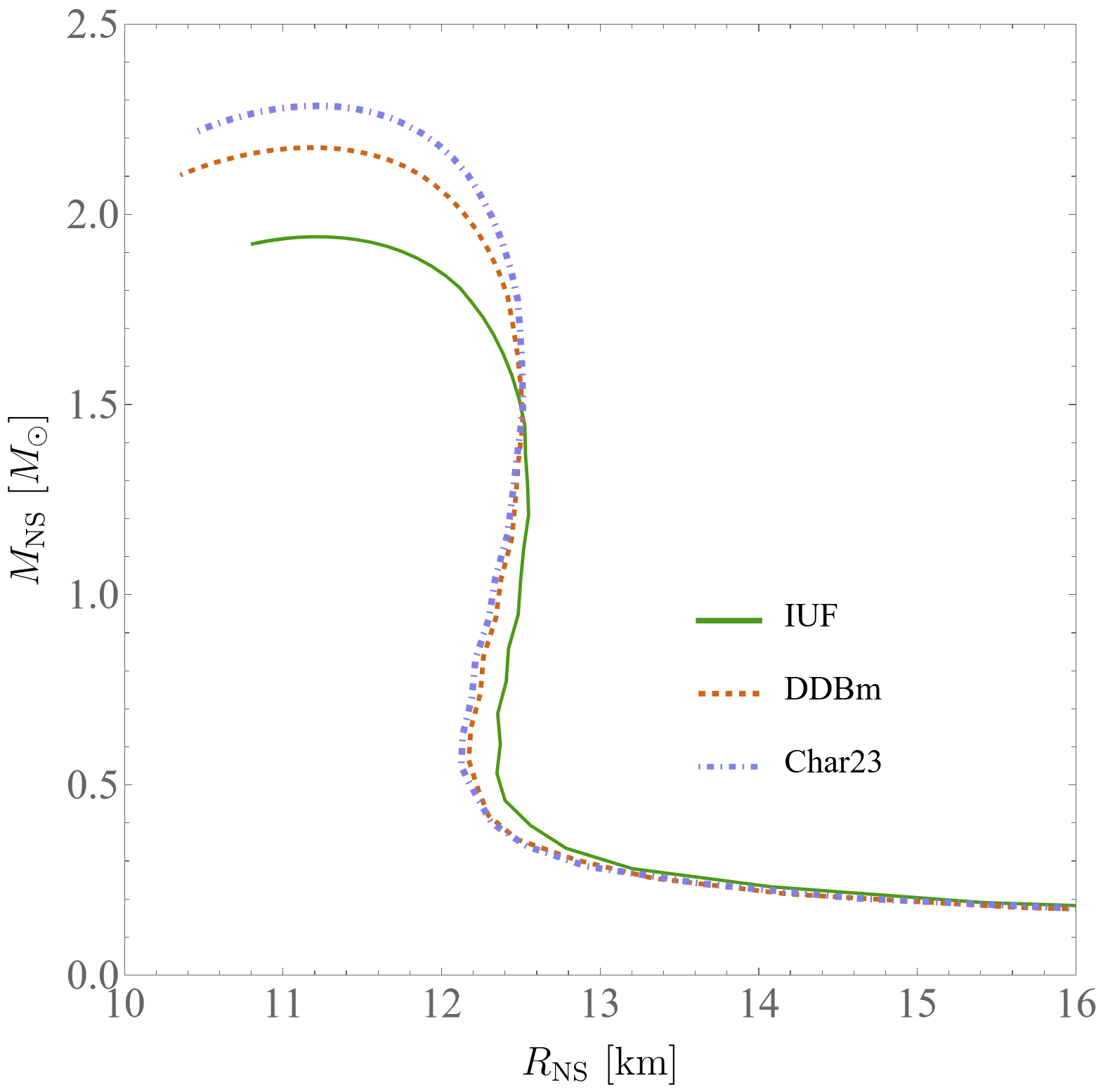}
  \caption{The EoS of J1856 solved with \texttt{CompactObject}. With a roughly constant radius around 12 km, large uncertainty in the mass translates to a large uncertainty in its density profile. We also plot the results from the DDBm model~\cite{Malik:2022zol} and model I from~\cite{Char:2023fue} (Char23 model). }
  \label{fig:EoS-J1856}
\end{figure}

Due to the large distance and extreme conditions of the NS core region, the axion emission rate from the core of J1856 succumbs to large systematic uncertainties. It is closely related to the benchmark values taken in Eq.~\eqref{eq:benchmark}. The main contributor is the mass uncertainty since a typical NS with a radius $r_0\simeq 12$~km is compatible with a wide range of mass around $(1-2) M_\odot$ (see, $e.g.$,~\cite{Steiner:2012xt}). Different masses profoundly affect the internal structure of NS, affecting axion emission via nucleon Fermi energy and general relativity effects. To estimate the size of the mass-induced uncertainty, we vary $M_{\rm NS}$ between 1.1 and 1.9 $ M_\odot$. To evaluate the axion production rate, we adopt the neutron bremsstrahlung formalism in~\cite{zotero-24826,Fortin:2018aom,Fortin:2021sst} with the single-pion exchange approximation. We solve the IUF~\cite{Fattoyev:2010mx} EoS for high-density nuclear matter with a customized~\texttt{CompactObject} package~\cite{Huang:2024rfg}. We then solve the Tolman-Oppenheimer-Volkoff equation for the relativistic celestial body with the EoS. The axion emission rate increases by more than 5 times when the NS mass changes from 1.1 to 1.9~$M_\odot$.
We show the EoS of J1856 in Fig.~\ref{fig:EoS-J1856}. By varying the mass from 1.1 to 1.9~$M_\odot$, we get the following luminosity in Table~\ref{tab:axion-lum-vs-ns-mass}, all normalized by the luminosity computed using the homogeneous density profile without taking into account of gravitational potential. We also plot two other EoS's as further references, namely the DDBm model described in~\cite{Malik:2022zol} and model I from~\cite{Char:2023fue} (Char23 model).
\begin{table}[b]
  \centering
  \begin{tabular}{c|ccccccccc}
    \hline\\
    $M_{\rm NS}$ [$M_\odot$] & 1.1 & 1.2 & 1.3 & \textbf{1.4} & 1.5 & 1.6 & 1.7 & 1.8 & 1.9 \\
    \hline \\
    $L_a$ [$L_{a,0}$] & 1.70 & 2.05 & 2.48 & \textbf{2.98} & 3.58 & 4.32 & 5.29 & 6.62 & 8.91 \\\\ $R_{\rm NS}$~[km] & 12.8 & 12.8 & 12.8 & 12.7 & 12.7 & 12.6 & 12.4 & 12.2 & 11.8 \\
    \hline
  \end{tabular}
  \caption{The axion luminosity assuming different masses for J1856, normalized by the luminosity assuming the homogeneous density profile without taking into account of gravitational potential, $L_{a,0}$. The benchmark is chosen as $M_{\rm NS} = 1.4 \, M_\odot$. The uncertainty in $L_a$ due to the mass profile is quantified in terms of the parameter $\kappa$ as shown by the gray band in Fig.~5 of the main text.}
  \label{tab:axion-lum-vs-ns-mass}
\end{table}


The subdominant contributor to the uncertainty is $D=123^{+11}_{-15}\,\mathrm{pc}$~\cite{Walter:2010ht}, where the impact is more straightforward.
One may also consider the uncertainty of the equatorial dipole $B$ field intensity $B_0=1.5\times 10^{13}$ G~\cite{vanKerkwijk:2007jp}. However, the uncertainty of the $B$ field is not directly available from the literature. Moreover, the $B$ field intensity only affects the final photon flux to the power of 0.4. Similarly, $T_{\rm SBB}^\infty$ and $R_{\rm SBB}$ are already tightly constrained by the X-ray spectrum below 1 keV, with relative uncertainties less than 2\%. We therefore ignore the uncertainties introduced by these three parameters safely. Consequently, the uncertainty range of $|g_{a\gamma}g_{aN}|^{2}$ is obtained by varying the above in parameters. As the prior of such uncertainties is not exactly known and unlikely to be Gaussian, we do not assign the range of $| g_{a\gamma}g_{aN}|$ with particular statistical interpretations. Note that instead of presenting the uncertainties in the contours from the fits, we show them as a band of the current constraint on the axion couplings by introducing the $\kappa$ parameter, which absorbs the uncertainties dominantly from the NS mass. $\kappa=1$ corresponds to our benchmark in Eq.~\eqref{eq:benchmark}. 
From the gray vertical band in Fig.~5 of the main text, it is clear that even though astrophysical and nuclear physics introduce large uncertainties in the axion emission and photon conversion rates, their impact on the coupling product is still limited as the coupling product is only sensitive to the inverse square root of the factors above.

\end{document}